\documentclass{article}

\usepackage{arxiv}

\usepackage[utf8]{inputenc} 
\usepackage[T1]{fontenc}    
\usepackage{hyperref}       
\usepackage{url}            
\usepackage{booktabs}       
\usepackage{amsfonts}       
\usepackage{nicefrac}       
\usepackage{microtype}      
\usepackage{lipsum}
\usepackage{graphicx}
\graphicspath{ {./images/} }
\usepackage{tikz}
    
\usetikzlibrary{calc}
\usepackage[framemethod=default]{mdframed}
 \usepackage{tabularx} 
 \usepackage{multirow}
 \usetikzlibrary{backgrounds,calc,shadings,shapes.arrows,shapes.symbols,shadows}

\tikzset{cvcv/.style={
     cloud, draw, aspect=2,color={black}
  }
}

\usepackage{pgfplots}
\usepackage{bchart}
\usepackage{pgfplotstable}
\pgfplotsset{compat=1.7}
\usepackage{amsmath}
\title{Performance Analysis of the Hybrid IoT Security Model of MQTT and UMA}

\author{ \href{https://orcid.org/0000-0000-0000-0000}{\includegraphics[scale=0.06]{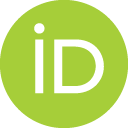}\hspace{1mm}Khalid S. Aloufi}\thanks{Use footnote for providing further
		information about author (webpage, alternative
		address)---\emph{not} for acknowledging funding agencies.} \\
	Department of Computer Engineering\\
	Taibah University\\
	Medina \\
	\texttt{koufi@taibahu.edu.sa} \\
	\And
	\href{https://orcid.org/0000-0000-0000-0000}{\includegraphics[scale=0.06]{orcid.png}\hspace{1mm}Omar H. Alhazmi} \\
	Department of Computer Science\\
	Taibah University\\
	Medina \\
	\texttt{ohhazmi@taibahu.edu.sa} \\
}

\begin{document}
\maketitle

\begin{abstract}
IoT applications are promising for future daily activities;
therefore, the number of IoT connected devices is expected to reach billions in the coming few years.
However, IoT has different application frameworks.
Furthermore, IoT applications require higher security standards.
In this work, an IoT application framework is presented with a security embedded structure using the integration between message queue telemetry transport (MQTT) and user managed access (UMA).
The performance analysis of the model is presented. 
Comparing the model with existing models and different design structures shows that the model presented in this work is promising for a functioning IoT design model with security.
The security in the model is a built-in feature in  its structure.
The model is built on recommended frameworks; therefore, it is ready for integration with other web standards for data sharing, which will help in making IoT application integrated from different developing parties.
\end{abstract}


\section{Introduction}
\label{sec:introduction}
Over the past few years, one of the remarkable emerging  technologies is the Internet of Things (IoT). It has vast varieties of possible applications, with a wide range of requirements in terms of security and performance. 
Several networking protocols are designed with performance in mind, then security is added later. Thus, security is often considered a burden and an overhead to the original protocol. The original non-secure protocol requires less communication and encryption overhead. Moreover, an IoT environment mainly consists of constrained devices, with limited resources and capabilities, which makes performance vital for the environment.
There is an essential need for a protocol that distributes the load and decentralizes security in order to have a scaleable IoT environment and satisfy the performance and security requirements at the same time.
One of the widely accepted IoT protocol is Message Queuing Telemetry Transport (MQTT). MQTT is an ISO standard (ISO/IEC PRF 20922)\cite{ISO}. It is a lightweight, publish-subscribe network protocol that transports messages between devices. MQTT typically runs over TCP/IP. It is designed for connections 
with remote locations where a "small code footprint" is required or the network bandwidth is limited. 
MQTT does not provide security by itself. Therefore, there have been some improvements to MQTT to improve its security; however, the overhead remains a critical issue in an environment with constrained devices. 

On the other hand, User Managed Access(UMA) provides security.
User-Managed Access (UMA) is an OAuth-based access management protocol standard. Version 1.0 of the standard was 
approved by the Kantara Initiative on March 23, 2015.\cite{KiS} 
OAuth is an open standard for access delegation, commonly used as a way for Internet users and applications to grant websites or applications access to their information on other websites. It leaves the power with the owner to control authorization.
It provides privacy and consent implications for web applications and the Internet of Things (IoT).

In the rest of this paper, we present some related work, then the background of MQTT and UMA; next, we propose a hybrid model by mapping MQTT and UMA. By simulation we evaluate the proposed secure MQTT/UMA hybrid system. Then the simulation experiment results is presented, discussed and analyzed; finally, a conclusion and final remarks are summarized and some future research directions are presented.

\section{Related Work} \label{Related Work}

Internet of Things deployment remains a major challenge; hence, many have suggested using the cloud
as a platform for the Internet of Things \cite{6803194}. It has also been  suggested that IoT sensors can be integrated into the cloud \cite{5599715}; moreover, the business model can be built around providing Sensing as a Service module (SenaaS) \cite{5678060}  

Neven Nikolov and Ognyan Nakov have suggested using MQTTS and SSL/SSH encryption for IoT devices\cite{8878636}. Such a scheme can be effective; however, the performance impact on IoT should be carefully investigated. Others have also proposed some amendments to MQTT by using Attribute Based Encryption (ABE) to secure some aspects of MQTT\cite{7280018}\cite{6883405}.

Perira et al. have listed challenges  of IoT, which are automated sensor configuration and context discovery  context acquisition, modeling, reasoning, and distribution   selection of sensors in  sensing-as-a-service  (SenaaS) models  security, and privacy-trust   and context sharing\cite{8769506}.

 Dai, Zheng and Zhang have proposed Block Chain of Things (BCoT) to secure IoT devices. Their model, seems very secure. However, the high demand for resources can make it infeasible with the current constrained IoT devices.
 \cite{8731639}
 
Niruntasukrat et al. have modified the OAuth 1.0a framework to accommodate the restrictions of IoT devices.\cite{7503802}

Ullah, Ahmed and Kim have proposed information-centric networking \cite{8555552}. They have elaborated on the importance of the fog layer, as a practical and gap filler layer in any deployment of the IoT/Cloud environment.

Aloufi and Alhazmi have shown how deploying IoT devices using fog computing has evident performance advantages 
\cite{6512846}. Moreover, in \cite{JJEE} they have suggested an IoT implementation by mapping MQTT protocols over the cloud-fog  by placing the broker on the fog; the results show that the proposed scheme improves performance and reliability.

Cruz-Piris et al. have proposed access control mechanism for IoT using Oauth 2.0 with MQTT\cite{Cruz}.

\section{Background}

\subsection{MQTT}\label{subsec:MQTT}
There are different IoT protocols, such as MQTT and CoAP. 
While CoAP is  highly competitive with MQTT. 
CoAP has a mechanism of exploration and observation; however, MQTT is much simpler to implement and much more popular \cite{6LoWPANAloufi}\cite{7845442}\cite{7814989}.
As shown in figure \ref{fig:MQTT}, MQTT consists of clients and a broker. The main unit is the broker, which manages messages and transactions between clients. Clients are either a subscriber or a publisher. The payload of messages transmitted between clients contains data, mainly a subject and its value. The publisher sends messages to the broker when it has an update message or a periodic message. The broker sends the messages to the subscribers of a specific subject.
Figure \ref{fig:publishsubscribeTransactions} shows that MQTT transaction.

\begin{figure*}[h]
\centering
\begin{tikzpicture}[ 
roundnode/.style={rectangle, draw=green!60, fill=green!5, very thick,dashed, minimum size=5mm,text width=1.8cm, text badly centered,minimum height=1cm},
squarednode/.style={rectangle, draw=red!60, fill=red!5, very thick,dashed, minimum size=5mm,text width=2cm, text badly centered,minimum height=1cm},
]

\coordinate (c) at (0,0);
\coordinate (d) at (0,2.4);
\coordinate (e) at (4,0);
\coordinate (f) at (4,2.4);
\coordinate (k) at (8,0);
\coordinate (l) at (8,2.4);

\node[squarednode] at (0,3) {$C_P$(Publishers)};
\node[squarednode] at (4,3) {MQTT};
\node[squarednode] at (8,3) {$C_S$(Subscribers)};

\draw 
 (c) -- (d) 
(e) -- (f) 
(k) -- (l);

\draw[-stealth] ($(c)!0.45!(d)$) -- node[scale=1,midway,fill=white]{$Connect$}($(e)!0.45!(f)$);
\draw[-stealth] ($(e)!0.25!(f)$) -- node[scale=1,midway,fill=white]{$Connection ACK$}($(c)!0.25!(d)$);
\draw[-stealth] ($(c)!0.05!(d)$) -- node[scale=1,midway,fill=white]{$publish(Topic,value)$}($(e)!0.05!(f)$);

\draw[-stealth] ($(e)!0.05!(f)$) -- node[scale=1,midway,fill=white]{$publish(Topic,value)$}($(k)!0.05!(l)$);

\draw[-stealth] ($(k)!0.95!(l)$) -- node[scale=1,midway,fill=white]{$Connect$}($(e)!0.95!(f)$);
\draw[-stealth] ($(e)!0.75!(f)$) -- node[scale=1,midway,fill=white]{$Connection ACK$}($(k)!0.75!(l)$);
\draw[-stealth] ($(k)!0.55!(l)$) -- node[scale=1,midway,fill=white]{$subscribe(Topic,value)$}($(e)!0.55!(f)$);
\end{tikzpicture}
\caption{MQTT publish and subscribe Transactions}
\label{fig:publishsubscribeTransactions}
\end{figure*}
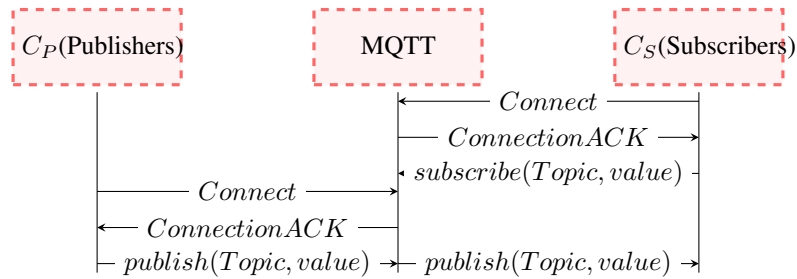

\begin{figure}[h]
\centering
\begin{mdframed}
\begin{tikzpicture}[
roundnode1/.style={circle, draw=green!60, fill=green!5, very thick,  minimum size=5mm,text width=1cm, text badly centered,minimum height=1cm},
roundnode2/.style={circle, draw=blue!60, fill=blue!5, very thick, minimum size=5mm,text width=1cm, text badly centered,minimum height=1cm},
squarednode/.style={rectangle, draw=red!60, fill=red!5, very thick,dashed, minimum size=5mm,text width=2cm, text badly centered,minimum height=2cm},
] 
 
\tikzstyle{squa2} = [draw, fill=blue!20,text width=4.5em, text badly centered,text width=2.8cm, minimum height=2cm]

\node[roundnode1]  (n1)  at (6,0) {MQTT \\Client};%
\node[roundnode1]  (n2)  at (6,-2) {MQTT \\Client};%
\node[squarednode]  (n3)  at (1,-2) {MQTT\\Broker};%
\node[roundnode1]  (n5)  at (6,-5) {MQTT \\Client};


\node[roundnode2]  (n7)  at (-3,0) {MQTT \\Client};%
\node[roundnode2]  (n8)  at (-3,-2) {MQTT \\Client};%
\node[roundnode2]  (n11)  at (-3,-5) {MQTT \\Client};

\node (ss1) at (-3,-3.5) {\vdots};
\node (ss2) at (6,-3.5) {\vdots};

\draw[dotted,->, very thick] (n1)  to[bend right=-0] node [scale=.8,midway,fill=white] {subscribe} (n3) ;
\draw[dotted,->, very thick] (n2)  to[bend right=0] node [scale=.8,midway,fill=white] {subscribe} (n3) ;
\draw[dotted,->, very thick] (n5)  to[bend right=0] node [scale=.8,midway,fill=white] {subscribe} (n3) ;

\draw[dotted,->, very thick] (n3)  to[bend right=-20] node [scale=.8,midway,fill=white] {message} (n1) ;
\draw[dotted,->, very thick] (n3)  to[out=10,in=160] node [scale=.8,midway,fill=white] {message} (n2) ;
\draw[dotted,->, very thick] (n3)  to[out=-20,in=120] node [scale=.8,midway,fill=white] {message} (n5) ;

\draw[dotted,->, very thick] (n7) -- (n3) node [scale=.8,midway,fill=white] (TextNode) {publish};
\draw[dotted,->, very thick] (n8) -- (n3) node [scale=.8,midway,fill=white] (TextNode) {publish};
\draw[dotted,->, very thick] (n11) -- (n3) node [scale=.8,midway,fill=white] (TextNode) {publish};
 
\end{tikzpicture}
\end{mdframed}
\caption{MQTT} 
\label{fig:MQTT}
\end{figure}

\begin{table}
\caption{Internet Of Things Stack}
\label{tab:mqtt}       
\begin{tabular}{|c|c|c|c|}
\hline\noalign{\smallskip}
Layer&OSI Layers&TCP/IP&IoT\\
\noalign{\smallskip}\hline\noalign{\smallskip}
7&Application&\multirow{3}{*}{Application}&\multirow{3}{*}{MQTT}\\
\cline{1-2}
6&Presentation &  &\\  \cline{1-2}
5&Session &  &\\\hline
4&Transport&Transport&TCP\\
\hline
\multirow{2}{*}{3}&\multirow{2}{*}{Network}&
\multirow{2}{*}{Internet}&\multirow{2}{*}{IP}\\ 
&&&\\
\hline
2&Data Link&\multirow{2}{*}{Network Access}&IEEE 802.15.4 MAC\\
\cline{1-2}
\cline{4-4}
1&Physical&&IEEE 802.15.4 PYS\\ 
\noalign{\smallskip}\hline
\end{tabular}
\end{table}

Table \ref{tab:mqtt} shows the IoT stack model for MQTT. With correspondence of the TCP/IP model, MQTT is one of the main IoT protocols because its messages are transmitted over TCP connection, in contrast to some other protocols,such as CoAP, which are transmitted over UDP. MQTT takes full advantage of the TCP/IP model, making application development and understanding of the protocol trivial. Also, MQTT integration over TCP help in the integration of MQTT with other protocols that share the same stack.

\subsection{UMA}\label{UMA}
OAuth2 is an open protocol to allow secure authorization of application without providing the password\cite{hardjono-oauth-umacore-13}.
User-Managed Access (UMA) is a profile of OAuth 2.0.
UMA defines how different entities communicates together.
UMA consists of  resource owner(RO),resource server(RS), authorization server(AS), client and the requesting party(RP).
The components work together to provide secure access to resources using protection API, authorization API  and tokens to access protected resource.

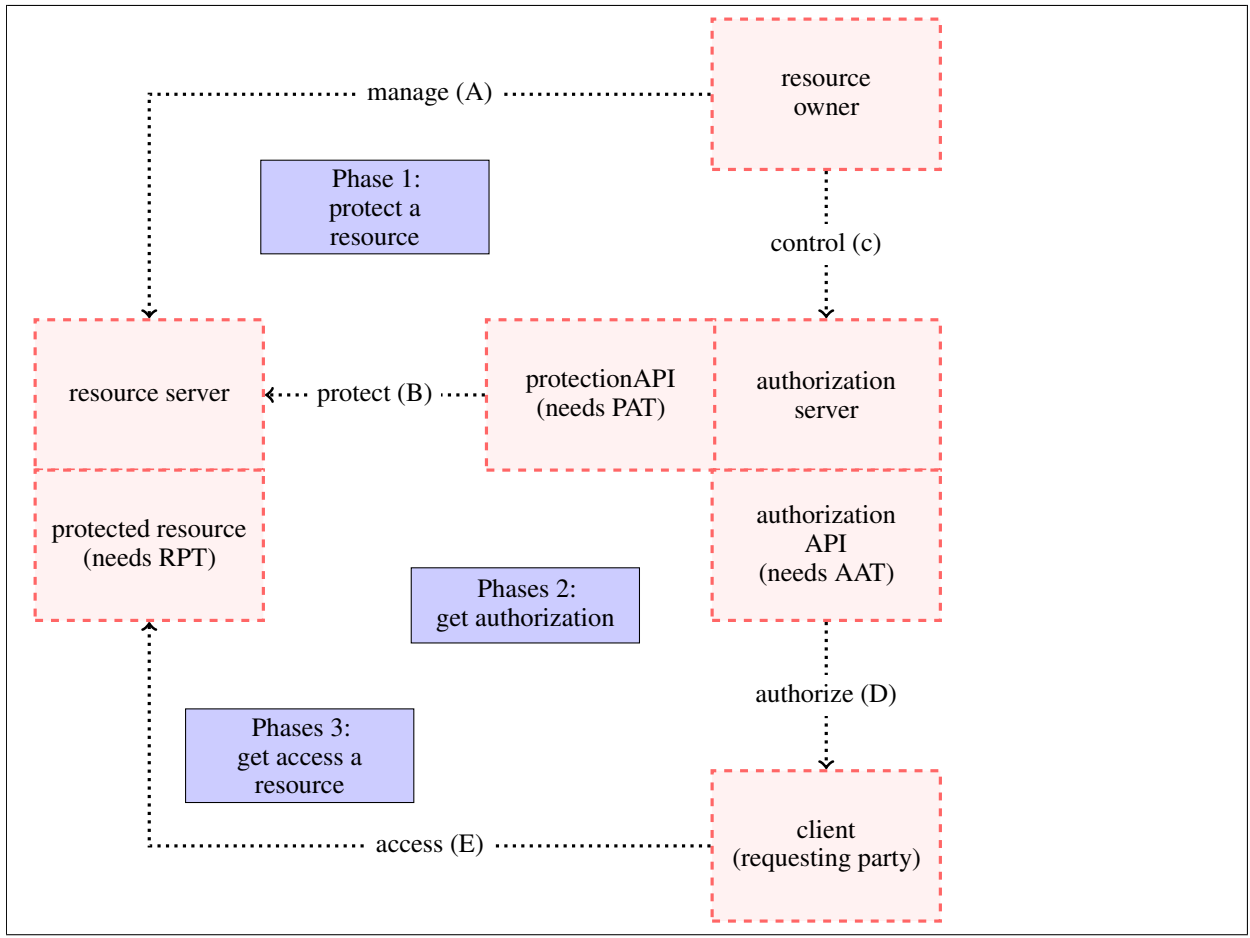
\begin{figure*}[h]
\centering
\begin{mdframed}
\begin{tikzpicture}[
roundnode/.style={circle, draw=green!60, fill=green!5, very thick, minimum size=7mm},
squarednode/.style={rectangle, draw=red!60, fill=red!5, very thick,dashed, minimum size=5mm,text width=2.8cm, text badly centered,minimum height=2cm},
] 
 
\tikzstyle{squa2} = [draw, fill=blue!20,text width=4.5em, text badly centered,text width=2.8cm, text badly centered,minimum height=1cm]
\tikzstyle{squa3} = [draw, fill=blue!20,text width=4.5em, text badly centered,text width=2.8cm, text badly centered,minimum height=1cm]

\node[squarednode]  (resourceowner)  at (6,1) {resource\\owner};
\node[squarednode]  (authorizationserver)  at (6,-3) {authorization \\server};
\node[squarednode]  (protectionAPIneedsPAT)  at (3,-3) {protectionAPI\\(needs PAT)};
\node[squarednode]  (authorizationAPIneedsAAT)  at (6,-5) {authorization\\API\\(needs AAT)};
\node[squarednode]  (client)  at (6,-9) {client\\(requesting party)};
\node[squarednode]  (resourceserver)  at (-3,-3) {resource server};
\node[squarednode]  (protectedresourceneedsRPT)  at (-3,-5) {protected resource (needs RPT)};

\draw[dotted,->, very thick,] (resourceowner) -- (authorizationserver) node [midway,fill=white] {control (c)};
\draw[dotted,->, very thick,] (authorizationAPIneedsAAT) --  (client) node [midway,fill=white] {authorize (D)};
\draw[dotted,->, very thick,] (protectionAPIneedsPAT) --  (resourceserver) node [midway,fill=white] {protect (B)};
c
\draw[dotted,->, very thick,] (resourceowner) -- (-3,1) node [midway,fill=white] {manage (A)} --  (resourceserver);
\draw[dotted,->, very thick,] (client) -- (-3,-9) node [midway,fill=white] {access (E)} --  (protectedresourceneedsRPT);

\node[squa2]  (Phase1)  at (0,-.5) {Phase 1: \\ protect   a \\ resource};
\node[squa2]  (Phases2)  at (2,-5.8) {Phases 2: \\get authorization};

\node[squa2]  (Phases3)  at (-1,-7.8) {Phases 3:\\get access a resource};


\end{tikzpicture}
\end{mdframed}
\caption{The Three Phases of the UMA Profile of OAuth,\cite{hardjono-oauth-umacore-13}.} 
\label{fig:The Three Phases of the UMA Profile of OAuth}
\end{figure*}
  
The resource owner controls resource  access by clients.
The client is operated by a requesting party.
Resources are hosted by a resource server.
An authorization server has roles defined by the resource owner, to access resources, .
For example, a resource owner can grant access to an application for one-time access to a resource following some roles defined by the owner and managed by the authorization server.

 There are three phases of the UMA of resources as shown in Figure \ref{fig:The Three Phases of the UMA Profile of OAuth}, which are protection, authorization and access\cite{hardjono-oauth-umacore-13}. 
 The figure shows the transaction of the three phases. In phase one, the resource owner is managing resources at the resource server, ("A"). The resource owner controls the authorization server,("C"), which provides the resources with protection API,(B). 
The protection API require a protection API token (PAT) to access a resource.
In phase two, the authorization phase, the client gets access to a set of resources in the resource server after being authorized by the authorization server . The authorization API is using an authorization API token (AAT) to get a requesting party token (RPT),("D").
Finally, in phase three, the client accesses a resource in the resource server using an RPT, ("E").

 Figure \ref{fig:SummarizedUMATransactions} shows the summarized transaction of the model. The client connects with the RS. 
 It gets the resource, then the client is connected with the AS to get grant access and finally the client can get the resource from the RS. This transaction is updated from the original one since the UMA model is updated and the client has no direct connection with the RO \cite{rfc6750}.

The UMA model has different APIs and tokens to be used.
Table \ref{tab:UMA API and Tokens} shows the list of APIs used, tokens, the issuer and the user of the tokens.
Using the protection API, a Protection API Token(PAT) is issued by the AS to the RS to access the protection API.
The RS is then issued using the RPT to protect a specific resource for a specific owner.
An Authorization API Token(AAT) is issued by the AS using the Authorization API when the RP contacts the AS with a well-formatted request as recommended by UMA \cite{hardjono-oauth-umacore-13}.
The RP will use the token to contact the Authorization API again to get a Requesting Party Token (RPT), which is the token that the RP is working to grant based on ATT.

Figure \ref{fig:DetailedUMATransactions} shows the detailed UMA transaction \cite{PrabathSiriwardena,maler-oauth-umagrant-00}. The RO logs in to the RS to share a R. Then, the RS connects with the AS to get a PAT for the R. The AS replies with a PAT with the R, RS and RO information. After that, when the client (RP) requests access to the R at the RS, the client requires the access credentials. For this reason, the client is communicating with the AS to get the RPT and AAT for the specific R. Before the access permission, the AS set the access permissions rules. Finally, the client gets the R with a valid RPT. The UMA data should be exchanged in JSON format\cite{PrabathSiriwardena}.
 
\begin{table}
\caption{UMA API and Tokens}
\label{tab:UMA API and Tokens}       
\begin{tabular}{|m{1.5cm}|m{1cm}|m{.8cm}|m{.8cm}|m{.8cm}|}
\hline\noalign{\smallskip}
Token&API&Issued By&Issed To&To Access\\
\noalign{\smallskip}\hline\noalign{\smallskip}

Authorization Party Token (AAT)&A API&AS&RP(Client)&A API\\\hline

protection API Token (PAT)&P API&AS&RS& P. API\\\hline

Requesting Party Token (RPT)&A API&AS&RP(Client)&R in RS\\

\noalign{\smallskip}\hline
\end{tabular}
\end{table}

\usetikzlibrary{calc}
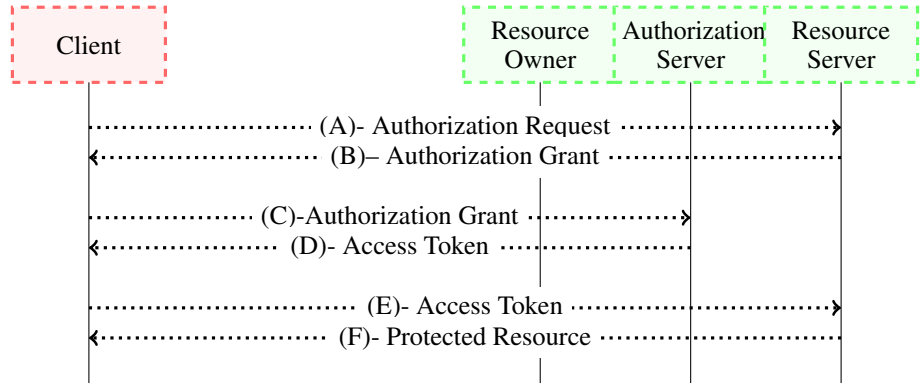
\begin{figure*}[h!]
\centering
\begin{tikzpicture}[
roundnode/.style={rectangle, draw=green!60, fill=green!5, very thick,dashed, minimum size=5mm,text width=1.8cm, text badly centered,minimum height=1cm},
squarednode/.style={rectangle, draw=red!60, fill=red!5, very thick,dashed, minimum size=5mm,text width=1.8cm, text badly centered,minimum height=1cm},
]

\node[squarednode] at (0,4.5) {Client};
\node[roundnode] at (6,4.5) {Resource \\ Owner};
\node[roundnode] at (8,4.5) {Authorization Server};
\node[roundnode] at (10,4.5) {Resource Server};

\coordinate (c) at (0,0);
\coordinate (d) at (0,4);
\coordinate (e) at (6,0);
\coordinate (f) at (6,4);
\coordinate (g) at (8,0);
\coordinate (h) at (8,4);
\coordinate (i) at (10,0);
\coordinate (j) at (10,4); 
\draw 
 (c) -- (d) 
(e) -- (f)  
(g) -- (h) 
(i) -- (j)  ;

\draw[dotted,->, very thick] ($(c)!0.85!(d)$) -- node[scale=1,midway,fill=white]{(A)- Authorization Request}($(i)!0.85!(j)$);
\draw[dotted,->, very thick] ($(i)!0.75!(j)$) -- node[scale=1,midway,fill=white]{(B)-- Authorization Grant}($(c)!0.75!(d)$);

\draw[dotted,->, very thick] ($(c)!0.55!(d)$) -- node[scale=1,midway,fill=white]{(C)-Authorization Grant}($(g)!0.55!(h)$);
\draw[dotted,->, very thick] ($(g)!0.45!(h)$) -- node[scale=1,midway,fill=white]{(D)- Access Token }($(c)!0.45!(d)$);

\draw[dotted,->, very thick] ($(c)!0.25!(d)$) -- node[scale=1,midway,fill=white]{(E)- Access Token}($(i)!0.25!(j)$);
\draw[dotted,->, very thick] ($(i)!0.15!(j)$) -- node[scale=1,midway,fill=white]{(F)- Protected Resource}($(c)!0.15!(d)$);

\end{tikzpicture}
\caption{Summarized UMA Transactions}
\label{fig:SummarizedUMATransactions}
\end{figure*}       

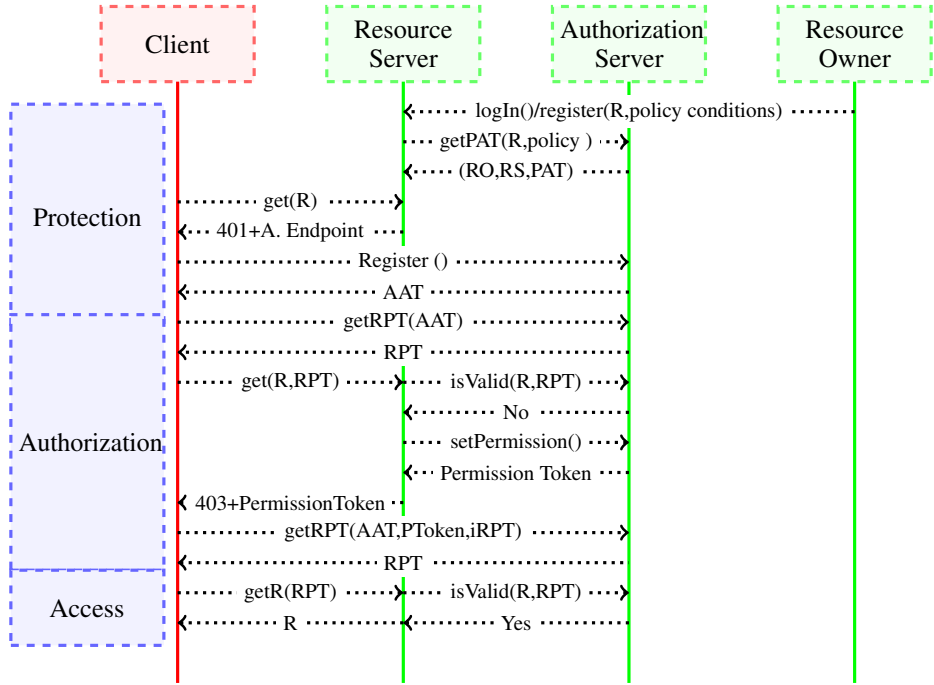
\begin{figure*}[h]
\centering
\begin{tikzpicture}[
roundnode/.style={rectangle, draw=green!60, fill=green!5, very thick,dashed, minimum size=5mm,text width=1.8cm, text badly centered,minimum height=1cm},
squarednode/.style={rectangle, draw=red!60, fill=red!5, very thick,dashed, minimum size=5mm,text width=1.8cm, text badly centered,minimum height=1cm},
] 

 \tikzstyle{roundnode4} = [rectangle, draw=blue!60, fill=blue!5, very thick,dashed, minimum size=5mm,text width=1.8cm, text badly centered,minimum height=3cm]
  \tikzstyle{roundnode5} = [rectangle, draw=blue!60, fill=blue!5, very thick,dashed, minimum size=5mm,text width=1.8cm, text badly centered,minimum height=3.4cm]
  
    \tikzstyle{roundnode6} = [rectangle, draw=blue!60, fill=blue!5, very thick,dashed, minimum size=5mm,text width=1.8cm, text badly centered,minimum height=1cm]

\node[squarednode] at (0,8.5) {Client};
\node[roundnode] at (9,8.5) {Resource Owner};
\node[roundnode] at (3,8.5) {Resource Server};
\node[roundnode] at (6,8.5) {Authorization Server};

\node[roundnode4] at (-1.2,6.2) {Protection};
\node[roundnode5] at (-1.2,3.2) {Authorization};
\node[roundnode6] at (-1.2,1) {Access};

\coordinate (c) at (0,0);
\coordinate (d) at (0,8);
\coordinate (g) at (3,0);
\coordinate (h) at (3,8);
\coordinate (e) at (6,0);
\coordinate (f) at (6,8);
\coordinate (i) at (9,0);
\coordinate (j) at (9,8); 
\draw[very thick,draw=red] 
 (c) -- (d); 
\draw [very thick, draw=green] (e) -- (f)  
(g) -- (h) 
(i) -- (j)  ;

\draw[dotted,->, very thick] ($(i)!0.95!(j)$) -- node[scale=0.8,midway,fill=white]{logIn()/register(R,policy conditions)}($(g)!0.95!(h)$);

\draw[dotted,->, very thick] ($(g)!0.90!(h)$) -- node[scale=0.8,midway,fill=white]{getPAT(R,policy )}($(e)!0.90!(f)$);

\draw[dotted,->, very thick] ($(e)!0.85!(f)$) -- node[scale=0.8,midway,fill=white]{(RO,RS,PAT)}($(g)!0.85!(h)$);

\draw[dotted,->, very thick] ($(c)!0.80!(d)$) -- node[scale=0.8,midway,fill=white]{get(R)}($(g)!0.80!(h)$);

\draw[dotted,->, very thick] ($(g)!0.75!(h)$) -- node[scale=0.8,midway,fill=white]{401+A. Endpoint}($(c)!0.75!(d)$);

\draw[dotted,->, very thick] ($(c)!0.70!(d)$) -- node[scale=0.8,midway,fill=white]{Register ()}($(e)!0.70!(f)$);

\draw[dotted,->, very thick] ($(e)!0.65!(f)$) -- node[scale=0.8,midway,fill=white]{AAT}($(c)!0.65!(d)$);

\draw[dotted,->, very thick] ($(c)!0.60!(d)$) -- node[scale=0.8,midway,fill=white]{getRPT(AAT)}($(e)!0.60!(f)$);

\draw[dotted,->, very thick] ($(e)!0.55!(f)$) -- node[scale=0.8,midway,fill=white]{RPT}($(c)!0.55!(d)$);

\draw[dotted,->, very thick] ($(c)!0.50!(d)$) -- node[scale=0.8,midway,fill=white]{get(R,RPT)}($(g)!0.50!(h)$);

\draw[dotted,<-, very thick] ($(e)!0.50!(f)$) -- node[scale=0.8,midway,fill=white]{isValid(R,RPT)}($(g)!0.50!(h)$);
\draw[dotted,->, very thick] ($(e)!0.45!(f)$) -- node[scale=0.8,midway,fill=white]{No}($(g)!0.45!(h)$);
\draw[dotted,<-, very thick] ($(e)!0.40!(f)$) -- node[scale=0.8,midway,fill=white]{setPermission()}($(g)!0.40!(h)$);
\draw[dotted,->, very thick] ($(e)!0.35!(f)$) -- node[scale=0.8,midway,fill=white]{Permission Token}($(g)!0.35!(h)$);
\draw[dotted,<-, very thick] ($(c)!0.30!(d)$) -- node[scale=0.8,midway,fill=white]{403+PermissionToken}($(g)!0.30!(h)$);

\draw[dotted,->, very thick] ($(c)!0.25!(d)$) -- node[scale=0.8,midway,fill=white]{getRPT(AAT,PToken,iRPT)}($(e)!0.25!(f)$);

\draw[dotted,<-, very thick] ($(c)!0.20!(d)$) -- node[scale=0.8,midway,fill=white]{RPT}($(e)!0.20!(f)$);

\draw[dotted,->, very thick] ($(c)!0.15!(d)$) -- node[scale=0.8,midway,fill=white]{getR(RPT)}($(g)!0.15!(h)$);

\draw[dotted,<-, very thick] ($(e)!0.15!(f)$) -- node[scale=0.8,midway,fill=white]{isValid(R,RPT)}($(g)!0.15!(h)$);

\draw[dotted,->, very thick] ($(e)!0.10!(f)$) -- node[scale=0.8,midway,fill=white]{Yes}($(g)!0.10!(h)$);

\draw[dotted,<-, very thick] ($(c)!0.10!(d)$) -- node[scale=0.8,midway,fill=white]{R}($(g)!0.10!(h)$);

\end{tikzpicture}
\caption{Detailed UMA Transactions}
\label{fig:DetailedUMATransactions}
\end{figure*}

\section{MQTT/UMA Hybrid Model}\label{sec:The Model Development}
The proposed model is a system composed of two known systems, which are the UMA and MQTT. In the following sections, the model is developed. The first step is to do a mapping between the functions of the MQTT and UMA protocols. Then, in the next step, the model is proposed. The proposed model considers both the functionality of MQTT and UMA to provide the features of both protocols. Therefore, the model extends the area of application of UMA to reach IoT devices that works with MQTT protocols.
The model proposed should increase the security level of small to large scale IoT application in smart city or smart building applications.

\subsection{UMA and MQTT Mapping}\label{sec:UMA and MQTT Mapping}
The model requires mapping between the functionality of both UMA and MQTT to work as one system. 
This section shows the mapping of the function of both protocols, UMA and MQTT.
Table \ref{tab:UMA and MQTT mapping} shows the mapping required for the model between MQTT and UMA to work effectively.
UMA consists of authorization server(AS),resource server(RS),resource owner(RO), client and requesting party (RP).
MQTT consists of MQTT broker,MQTT publisher and MQTT subscribers.
The AS,RS,RO, client and RP are mapped with the MQTT broker,MQTT publisher and MQTT subscribers.
The model, which will be detailed in section \ref{sec:themodel}, consists of one UMA and two MQTT.
MQTT manages topics which is considered a R in UMA.

The first MQTT consists of MQTT Broker 1 (MB1), subscriber (S1) and publisher (P1).
The second MQTT consists of MQTT Broker 2 (MB1), subscriber (S2) and publisher (P2). 
The model works by having P1 publish topics from IoT devices.
MB1 gets the published topics from P1.
After-that, MB1 publishes topics to MB2.
 S1 is MB2 and P2 is MB1.
After the topic is published from P1 to MB1, MB2 gets the published topics from MB1.
Finally, MB2 publishes topics to S2, the subscriber in the second MQTT.

To add UMA to the model, MB1 is mapped with the RO because of the similar functionality of both since both have resources to share.
The S2 is mapped with a RP since S2 subscribes to a topic with the MB2.
The S2 is considered as a RP because it will require use of a client to access a resource published by a MB2.
The MB2 works as a RS because of the similar functionality of holding the resources access permissions.
For the model to be practical and logical for implementations, the AS and client are considered two different standalone fog servers.

\begin{table}
\centering
\caption{Mapping  UMA and MQTT }
\label{tab:UMA and MQTT mapping}       
\begin{tabular}{|c|c|}
\hline\noalign{\smallskip}
MQTT&UMA\\
\noalign{\smallskip}\hline\noalign{\smallskip}
P1       &non\\\hline
MB1/P2    &RO\\\hline
S1/MB2       &RS\\\hline
non                 &AS\\\hline
S2      &RP\\\hline
non                 &Client\\\hline
topic/subject                 &R\\

\noalign{\smallskip}\hline
\end{tabular}
\end{table}

One of the main transactions added by the MQTT broker is the publishing of any new topic value.
Now the RS has the resource and has its value.
The resource is available for any successful RP access.
However, when a new value of a resource is received, RS has to update the resource and publish the resource to its subscribers.
The integration between the functionality of the MB2 and RS is the main addition of the model.

\subsection{The Model}\label{sec:themodel}
\begin{figure}[h]

\begin{tikzpicture}[every node/.style={circle, fill=green!5,minimum width=5mm,draw,shading=axis,top color=gray!20}]

\tikzstyle{roundnode11} = [rectangle, draw=blue!60, fill=blue!50, very thick,dashed, minimum size=5mm,text width=8.25cm,minimum height=3cm] 
\tikzstyle{roundnode16} = [rotate=90,rectangle, draw=blue!60, fill=blue!50, very thick,dashed,] 

\node[roundnode11] at (4.5,4) {};
\node[roundnode16] at (-.1,4) {Fog Layer};


\node (a) at (0,0)     {P1};
\node (b) at (0,1)     {P1};
\node (c) at (0,2)     {P1}; 
\node[cvcv,scale=2]  (d2) at (5,7)    {Internet}; 
\node[text width=1.5cm,align=center,font=\small ] (d) at (2,1) {MB1/P2/RO}; 
\node[cvcv,text width=1cm,align=center,font=\small ] (e) at (8,4) {AS}; 
\node[cvcv,text width=2cm,align=center,font=\small ] (e3) at (2,4) {S1/MB2/RS}; 
\node[text width=1cm,align=center,font=\small] (i3) at (5,1) {S2/RP}; 

\node[cvcv,text width=1cm,align=center,font=\small ] (g) at (5,4)     {Client}; 


\draw (a) -- (d);
\draw (b) -- (d);
\draw (c) -- (d);
\draw (e) -- (g);
\draw (g) -- (i3); 
\draw (d) -- (e3); 
\draw (d2) -- (e); 
\draw (e3) -- (d2);
\draw (g) -- (d2);
\draw (e3) -- (d2);
\draw (e3) -- (g);

\end{tikzpicture}
      \caption{System Model}
\label{fig:networkmodel}

\end{figure}
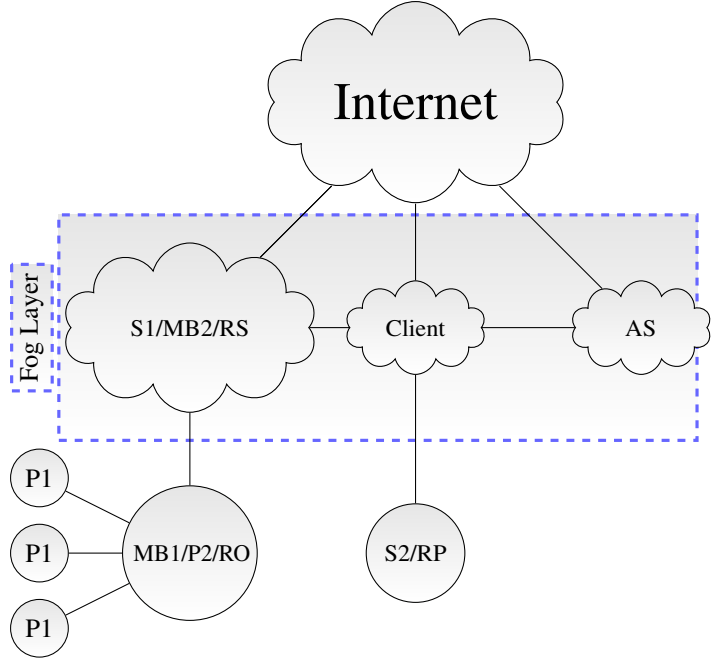

Figure \ref{fig:networkmodel} shows the applicable network model proposed in this work.
The model consists of the P1 as a publisher, S2/RP as a subscriber, RS/MB2/S1, MB1/P2/RO, client and AS.
As will be shown later, most of the transaction messages of the proposed model are between the three main entities, which are S1/MB2/RS, client and AS.
Therefore, to increase the performance of the model, RS/MB2/S1, Client and AS are located  in the fog layer.

As known and shown earlier in section \ref{subsec:MQTT}, the basic MQTT model consists of a MQTT broker, subscribers and publishers.
In the model two MQTT models are joined to make a new MQTT model.
Technically there are no P2 and S1 in the system. P2 and S1 functionality are integrated in MB1 and MB2, consequently.

S2/RP gets the messages from the fog layer, which is expected to provide more connectivity in availability and connection speed compared to the first mode as well as more security enhancement with UMA model.
MB2 is located in the fog layer, which is connected directly with MB1.

The model can be extended by having the MB2 connected to more than one MB1.
Furthermore, each MB1 could be using one or more methods for connectivity to the fog layer.
The fog layer provides more performance and security of the high Quality of Service (QoS) required of a large-scale application \cite{Alkhafajiy}.
One of the expected benefits of using the fog cloud is to decrease the response time between the subscribers and publishers. 
The publishers notify the broker for any updates and the broker notifies the subscribers. The broker also has the log of each publisher in case it is needed by subscribers or for more statistical computations, such as the average value of a specific publisher or timing values, such as the last time a specific publisher updated its value.

\section{Model Evaluation}

In this section, we describe the evaluation methodology, which consists of four parts: the simulation setup,  the simulation presentation; the results and discussion are presented afterward. 

\subsection{System Initial Configuration}
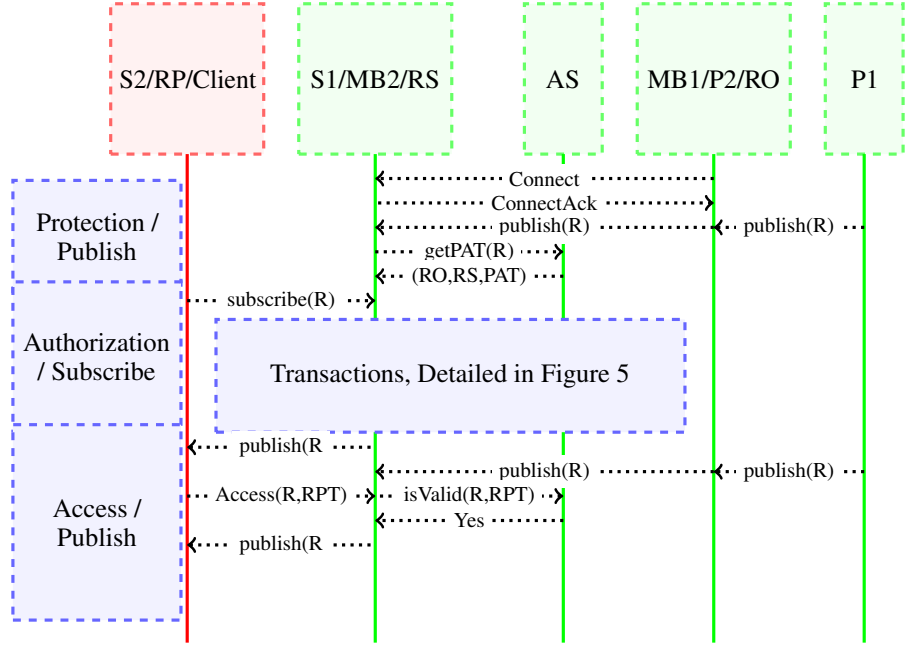
\begin{figure*}[h]
\centering
\begin{tikzpicture}[
roundnode/.style={rectangle, draw=green!60, fill=green!5, very thick,dashed, minimum size=5mm,text width=1.8cm, text badly centered,minimum height=2cm},
squarednode/.style={rectangle, draw=red!60, fill=red!5, very thick,dashed, minimum size=5mm,text width=1.8cm, text badly centered,minimum height=2cm},
] 

\tikzstyle{roundnode4} = [rectangle, draw=blue!60, fill=blue!5, very thick,dashed, minimum size=5mm,text width=1.8cm, text badly centered,minimum height=3cm]
\tikzstyle{roundnode5} = [rectangle, draw=blue!60, fill=blue!5, very thick,dashed, minimum size=5mm,text width=5cm, text badly centered,minimum height=2cm]
  
\tikzstyle{roundnode6} = [rectangle, draw=blue!60, fill=blue!5, very thick,dashed, minimum size=5mm,text width=1.8cm, text badly centered,minimum height=1cm]
\tikzstyle{roundnode7} = [rectangle, draw=blue!60, fill=blue!5, very thick,dashed, minimum size=5mm,text width=2cm, text badly centered,minimum height=1.5cm]  
\tikzstyle{roundnode8} = [rectangle, draw=blue!60, fill=blue!5, very thick,dashed, minimum size=5mm,text width=2cm, text badly centered,minimum height=2.6cm]  
\tikzstyle{roundnode9} = [rectangle, draw=blue!60, fill=blue!5, very thick,dashed, minimum size=5mm,text width=2cm, text badly centered,minimum height=2cm] 
\tikzstyle{roundnode11} = [rectangle, draw=blue!60, fill=blue!5, very thick,dashed, minimum size=5mm,text width=6cm, text badly centered,minimum height=1.5cm] 
 
 \tikzstyle{roundnode12} = [rectangle, draw=green!60, fill=green!5, very thick,dashed, minimum size=5mm,text width=.8cm, text badly centered,minimum height=2cm]

\node[squarednode] at (0,4.5) {S2/RP/Client};
\node[roundnode] at (7,4.5) {MB1/P2/RO};
\node[roundnode12] at (9,4.5) {P1};
\node[roundnode] at (2.5,4.5) {S1/MB2/RS};
\node[roundnode12] at (5,4.5) {AS};

\coordinate (c) at (0,-3);
\coordinate (d) at (0,3.5);
\coordinate (g) at (2.5,-3);
\coordinate (h) at (2.5,3.5);
\coordinate (e) at (5,-3);
\coordinate (f) at (5,3.5);
\coordinate (i) at (7,-3);
\coordinate (j) at (7,3.5); 
\coordinate (k) at (9,-3);
\coordinate (l) at (9,3.5); 

\draw[very thick,draw=red] 
 (c) -- (d); 
\draw [very thick, draw=green] 
(e) -- (f)  
(g) -- (h) 
(i) -- (j)
(k) -- (l);

\node[roundnode7] at (-1.2,2.4) {Protection / Publish};
\node[roundnode9] at (-1.2,.8) {Authorization / Subscribe};
\node[roundnode11] at (3.5,.55) {Transactions, Detailed in Figure  \ref{fig:DetailedUMATransactions}};
\node[roundnode8] at (-1.2,-1.4) {Access / Publish};

\draw[dotted,->, very thick] ($(c)!0.70!(d)$) -- node[scale=0.8,midway,fill=white]{subscribe(R)}($(g)!0.70!(h)$);


 \draw[dotted,<-, very thick] ($(c)!0.40!(d)$) -- node[scale=0.8,midway,fill=white]{publish(R}($(g)!0.40!(h)$);
 
 \draw[dotted,<-, very thick] ($(c)!0.20!(d)$) -- node[scale=0.8,midway,fill=white]{publish(R}($(g)!0.20!(h)$);

Fig. 7: MQTT UMA Subscribe Transactions
\draw[dotted,->, very thick] ($(i)!0.95!(j)$) -- node[scale=0.8,midway,fill=white]{Connect}($(g)!0.95!(h)$);
\draw[dotted,<-, very thick] ($(i)!0.90!(j)$) -- node[scale=0.8,midway,fill=white]{ConnectAck}($(g)!0.90!(h)$);
\draw[dotted,->, very thick] ($(i)!0.85!(j)$) -- node[scale=0.8,midway,fill=white]{publish(R)}($(g)!0.85!(h)$);
\draw[dotted,->, very thick] ($(k)!0.85!(l)$) -- node[scale=0.8,midway,fill=white]{publish(R)}($(i)!0.85!(j)$);
\draw[dotted,->, very thick] ($(g)!0.80!(h)$) -- node[scale=0.8,midway,fill=white]{getPAT(R)}($(e)!0.80!(f)$);

\draw[dotted,->, very thick] ($(e)!0.75!(f)$) -- node[scale=0.8,midway,fill=white]{(RO,RS,PAT)}($(g)!0.75!(h)$);

\draw[dotted,->, very thick] ($(i)!0.35!(j)$) -- node[scale=0.8,midway,fill=white]{publish(R)}($(g)!0.35!(h)$);

\draw[dotted,->, very thick] ($(k)!0.35!(l)$) -- node[scale=0.8,midway,fill=white]{publish(R)}($(i)!0.35!(j)$);

\draw[dotted,->, very thick] ($(c)!0.30!(d)$) -- node[scale=0.8,midway,fill=white]{Access(R,RPT)}($(g)!0.30!(h)$);
\draw[dotted,<-, very thick] ($(e)!0.30!(f)$) -- node[scale=0.8,midway,fill=white]{isValid(R,RPT)}($(g)!0.30!(h)$);

\draw[dotted,->, very thick] ($(e)!0.25!(f)$) -- node[scale=0.8,midway,fill=white]{Yes}($(g)!0.25!(h)$);

\end{tikzpicture}
\caption{MQTT UMA Subscribe Transactions}
\label{fig:MQTT UMA Subscribe Transactions}
\end{figure*}

This section shows the results and discussion of the simulation experiments of the IoT model shown earlier in section \ref{sec:themodel}.
As shown in figure \ref{fig:MQTT UMA Subscribe Transactions}, the  model contains P1, MB1/P2/RO, S1/MB2/RS, client, S2/RP and  AS. 

While P1 is in direct connection to MB1/P2/RO, there is no direct connection between MB1/P2/RO and S2/RP.
S2/RP gets the published topic from S1/MB2/RS, which is well established in security and performance in the fog layer.
MB1/P2/RO is expected to be a private property node in a home or a building at a security level that should not establish connection with general connections, such as RPs.
Also, S1/MB2/RS is assumed to be connected with  MB1/P2/RO with normal Internet connection. 

This is an increase in security without added security overhead for MB1/ P2/RO messages.
Effectively, the gained security overhead for messages is added to messages between S1/MB2/RS and MB1/P2/RO and between  S1/MB2/RS and S2/RP.

In section \ref{UMA}, Figure \ref{fig:DetailedUMATransactions} shows the detailed UMA transactions to add security for general systems.
Figure \ref{fig:MQTT UMA Subscribe Transactions}
shows messages exchange between UMA and MQTT units to provide a reliable IoT system that uses UMA as an added security layer.
Most messages transactions in the model are between the RS, client and AS. 
Therefore, S1/MB2/RS , AS and client are allocated in the model in the fog layer with high speed connection with each other to increase QoS.

\subsection{Evaluating the Model}\label{sec:experiments}
The model evaluation is done through simulation of the MQTT/UMA hybrid system.
To configure the simulation model, the system is configured as follows.
The wireless connection with any entity in the fog layer is 120 Mbps, assumed by a ping to the MQTT broker at test.mosquitto.org, and also tested by Jaloudi \cite{Jaloudi}. 
MB1/P2/RO can get from 10-100 Mbps over 802.11g protocol or wired connection. 
Therefore, MB1/P2/RO can connect with several IoT devices at once, with a high data rate from IoT.
However, the sending rate of each IoT device is much lower, allowing the broker to receive data from several IoT devices. 
Each IoT device is equipped with Zigbee, using an IEEE 802.15.4 antenna, with a rate of 250 kbps and the maximum message size is 127 bytes, according to the Zigbee Specification  \cite{10.1007/978-3-540-85500-2_26}.
As a result, the transmission time of one message is about 4 ms.

Fog servers are assumed to be equipped with powerful nodes, such as Intel Xeon W-3275M @ 2.50GHz.
In the fog server, the mean time required to access a database at the S1/MB1/RS  is 13 ms
The mean time required to query the database is about 10 ms.
The mean processing time at  the S1/MB2/RS, AS, and client is 10 ms.
IoT devices represented by $P1$ are publishing data with the Poisson arrival process.

A ping between two servers in Europe, assumed to be servers in the fog layer, requires between 3 ms and 100 ms, depending on the location of the fog servers. In this work, it is assumed that the fog server is in the same area, such as a country. Therefore, the assumed time is about 3 ms to exchange a message between two fog servers.
The tested ping between two fog servers in different areas, like the US and Europe, is about 200 ms.
From  the simulation experiments,  the arrival rate of data from each of the IoT devices is 127 bytes per second, which is one reading of a subject data keeping the data size minimal to avoid fragmentation.

\begin{table}
\caption{Processing and transaction required time}
\label{tab:Processing and transaction required time}       
\centering
\begin{tabular}{|l|l|}

\hline\noalign{\smallskip}
Parameter&$t$ (ms)\\
\noalign{\smallskip}\hline\noalign{\smallskip}

$T_{P1}$& 10 \\\hline

$T_{MB1/P2/RO}$&100 \\\hline

$T_{AS}$&13  +10  +10 \\\hline

$T_{S1/MB2/RS}$&13  +10  +10 \\\hline

$T_{S2/RP}$&100 \\\hline

$T_{Client}$&10 \\\hline

$T_{P1xMB1/P2/RO}$&4 \\\hline

$T_{MB1/P2/ROxS1/MB2/RS}$& 200 \\\hline

$T_{S1/MB2/RSxAS}$& 3 \\\hline

$T_{ClientxS1/MB2/RS}$& 3 \\\hline

$T_{ClientxAS}$& 3 \\\hline

$T_{S2/RPxClient}$& 200 \\ 

\noalign{\smallskip}\hline
\end{tabular}
\end{table}

\begin{figure}[h]
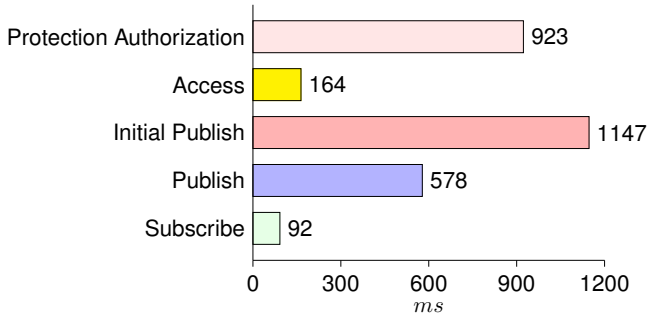

\scalebox{.85}{
\begin{bchart}[step=300,max=1200,width=5.5cm]
\bcbar[label= Protection Authorization,color=red!10]{923}
\smallskip
\bcbar[label= Access,color=yellow]{164}
\smallskip
\bcbar[label=Initial Publish,color=red!30]{1147}
\smallskip
\bcbar[label=Publish,color=blue!30]{578}
\smallskip
\bcbar[label=Subscribe ,color=green!10]{92}
\bcxlabel{$ms$}
\end{bchart}}
\caption{Phases Processing Time} 
\label{fig:Phases Processing Time}
\end{figure}

Table \ref{tab:Processing and transaction required time} shows the transaction time between the different nodes of the model,
and the processing time at each node.
The table is the conclusion of the configuration of the model.
The table is used to configure the simulation.
Processing time at the IoT device is defined by $T_{P1}$.
The table symbols are shown in Figure \ref{fig:Phases Processing Time}.

\begin{figure}
\centering
\begin{tikzpicture}
\begin{axis}[xlabel=$\lambda$,ylabel=System Utilization,legend pos=north west]
\addplot [mark=*, blue] coordinates { 
(0.0017006802721088435,0.9982993197278911)
(0.0016983695652173915,0.9969429347826088)
(0.00169606512890095,0.9955902306648577)
(0.001693766937669377,0.9942411924119243)
(0.0016914749661705011,0.9928958051420842)
(0.00168918918918919,0.9915540540540545)
(0.0016869095816464245,0.9902159244264511)
(0.0016846361185983837,0.9888814016172512)
(0.0016823687752355327,0.9875504710632577)
(0.0016801075268817215,0.9862231182795705)
(0.0016778523489932899,0.9848993288590612)
(0.0016756032171581783,0.9835790884718506)
(0.0016733601070950484,0.9822623828647934)
(0.0016711229946524081,0.9809491978609636)
(0.0016688918558077455,0.9796395193591466)
(0.0016666666666666685,0.9783333333333344)
(0.0016644474034620527,0.9770306258322249)
(0.0016622340425531937,0.9757313829787246)
(0.001660026560424969,0.9744355909694568)
(0.0016578249336870051,0.973143236074272)
(0.0016556291390728503,0.971854304635763)
(0.001653439153439156,0.9705687830687846)
(0.001651254953764864,0.9692866578599751)
(0.0016490765171503986,0.9680079155672839)
(0.0016469038208168673,0.966732542819501)
(0.0016447368421052663,0.9654605263157913)
(0.001642575558475693,0.9641918528252318)
(0.001640419947506565,0.9629265091863536)
(0.0016382699868938435,0.9616644823066861)
(0.0016361256544502652,0.9604057591623056)
(0.001633986928104579,0.9591503267973878)
(0.001631853785900787,0.9578981723237621)
(0.0016297262059973964,0.9566492829204717)
(0.0016276041666666706,0.9554036458333357)
(0.0016254876462938922,0.9541612483745148)
(0.0016233766233766276,0.9529220779220804)
(0.001621271076523999,0.9516861219195875)
(0.001619170984455963,0.9504533678756503)
(0.0016170763260025917,0.9492238033635213)
(0.0016149870801033638,0.9479974160206746)
(0.0016129032258064564,0.9467741935483899)
(0.0016108247422680461,0.9455541237113431)
(0.0016087516087516136,0.9443371943371972)
(0.0016066838046272544,0.9431233933161983)
(0.0016046213093709935,0.9419127086007731)
(0.0016025641025641077,0.9407051282051312)
(0.001600512163892451,0.9395006402048687)
(0.0015984654731457854,0.938299232736576)
(0.0015964240102171192,0.937100893997449)
(0.0015943877551020465,0.9359056122449013)
(0.0015923566878980949,0.9347133757961816)
(0.0015903307888040772,0.9335241730279933)
(0.001588310038119447,0.9323379923761154)
(0.0015862944162436609,0.9311548223350289)
(0.001584283903675545,0.9299746514575449)
(0.0015822784810126645,0.928797468354434)
(0.0015802781289507017,0.927623261694062)
(0.0015782828282828348,0.926452020202024)
(0.0015762925598991238,0.9252837326607857)
(0.0015743073047859008,0.9241183879093238)
(0.001572327044025164,0.9229559748427713)
(0.0015703517587939766,0.9217964824120642)
(0.0015683814303638715,0.9206398996235925)
(0.0015664160401002577,0.9194862155388512)
(0.0015644555694618344,0.9183354192740968)
(0.0015625000000000072,0.9171875000000043)
};
\end{axis}
\end{tikzpicture}
\caption{System Utilization}
\label{fig:System Utilization}
\end{figure}
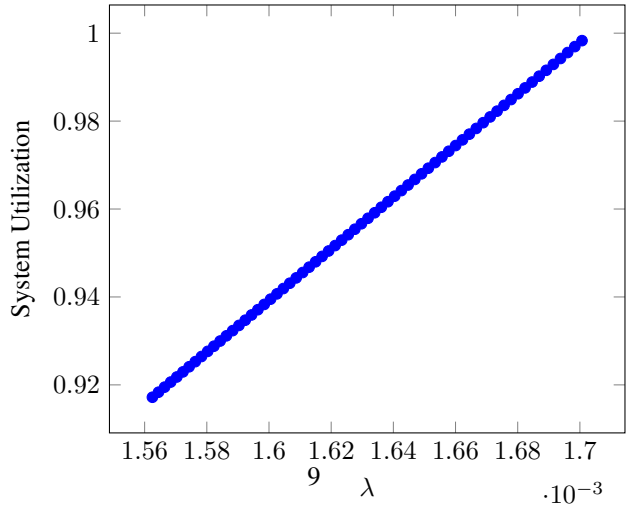

The transmission time between IoT device $P1$  and MQTT Broker $MB1/P2/RO$ is defined by $T_{P1xMB1/P2/RO}$. The same applies to other transactions. The symbols shown in the table are as follows.
$P1$  is the IoT devices.
$MB1$ is the first broker connected directly with $P1$ and also connected with the $S1$ node. as the subscriber.
$MB1$ shares the node with $P2$ and $RO$. The model is mutually inclusive between the two-broker system of $MB1$ and $MB2$.
$P2$ is the publisher of the second broker system. As mentioned, $P2$ share the node with $MB1$ and $RO$.
$MB2$ is the second broker and  share the node with $S1$ and $RS$.
$MB2$ is connected with $S2$ through the $Client$.
$S2$ shares its node with $RP$.

\begin{equation}
 \begin{aligned}
 \label{ProtectionAutherization}
 \begin{split}
1xT_{MB1/P2/RO}+
13xT_{AS}+\\
10xT_{S1/MB2/RS}+\\
10xT_{Client}+\\
1xT_{MB1/P2/ROxS1/MB2/RS}+\\
6xT_{S1/MB2/RSxAS}+\\
4xT_{ClientxS1/MB2/RS}+\\
6xT_{ClientxAS}
&=1207  ms
\end{split}
\end{aligned}
\end{equation}

\begin{equation}
 \begin{aligned}
 \label{Access}
 \begin{split}
2xT_{AS}+
2xT_{S1/MB2/RS}+\\
2xT_{Client}+\\
2xT_{S1/MB2/RSxAS}+\\
2xT_{ClientxS1/MB2/RS} 
&= 164 ms
\end{split}
\end{aligned}
\end{equation}

\begin{equation}
 \begin{aligned}
 \label{InitialPublish}
 \begin{split}
1xT_{P1}+
3xT_{MB1/P2/RO}+\\
2xT_{AS}+
5xT_{S1/MB2/RS}+\\
3xT_{MB1/P2/ROxS1/MB2/RS}+\\
2xT_{S1/MB2/RSxAS}
&= 1147 ms
\end{split}
\end{aligned}
\end{equation}

\begin{equation}
\label{Publish}
\begin{aligned}
\begin{split}
1xT_{P1}+
1xT_{MB1/P2/RO}+\\
2xT_{AS}+\\
2xT_{S1/MB2/RS}+\\
1xT_{S2/RP}+\\
2xT_{Client}+
1xT_{P1xMB1/P2/RO}+\\
1xT_{MB1/P2/ROxS1/MB2/RS}+\\
2xT_{S1/MB2/RSxAS}+\\
2xT_{ClientxS1/MB2/RS}\\
&=578 ms
\end{split}
\end{aligned}
\end{equation}

\begin{equation}
 \begin{aligned}
\label{Subscribe}
\begin{split}
2xT_{S1/MB2/RS}+\\
2xT_{Client}+\\
2xT_{ClientxS1/MB2/RS}
&=92 ms
\end{split}
\end{aligned}
\end{equation}

Equation \ref{ProtectionAutherization} shows that 1207 $ms$ is required for protection and authorization, .
For initial access, Equation \ref{Access} shows that 164 $ms$ is required.
Equation \ref{InitialPublish} shows the  time of initial publishing is 1147 $ms$.
For subsequent publishing, the time required is 578 $ms$ as shown by Equation \ref{Publish}.
In Figure \ref{fig:MQTT UMA Subscribe Transactions}, the function $Access(R,RPT)$ is invoked only when the client initiates the request; otherwise, the published message reaches to the $RP$ as a subscriber to the subject.
 Equation \ref{Subscribe} shows the subscriber transaction time is 92 $ms$.

\begin{figure}
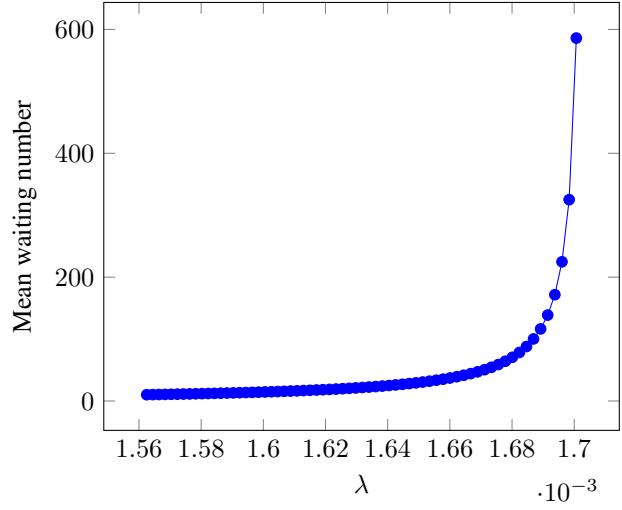

\centering

\caption{Mean waiting number}
\label{fig:Mean waiting number}
\end{figure}

\begin{figure}
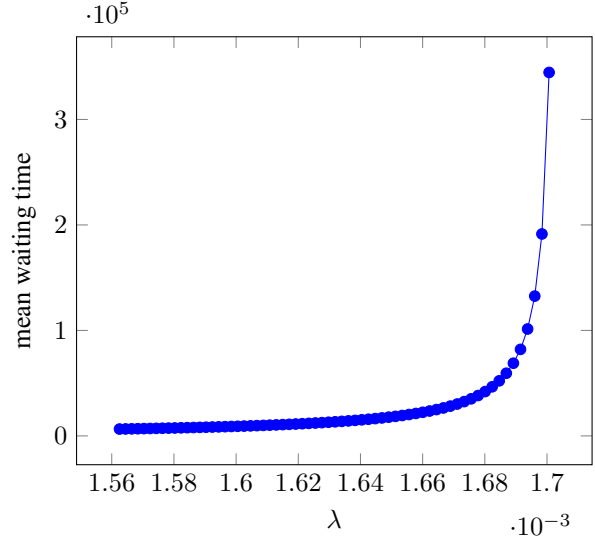

\centering
%
\caption{mean waiting time}
\label{fig:mean waiting time}
\end{figure}

\begin{figure}
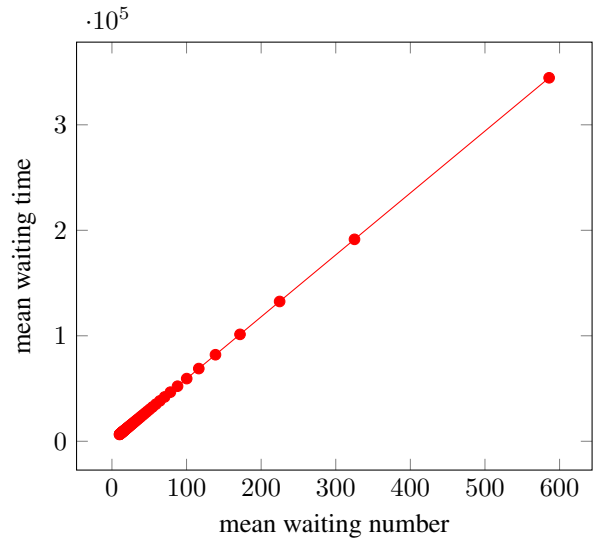

\centering
%
\caption{mean waiting time vs  mean waiting number}
\label{fig:mean waiting time vs  mean waiting number}
\end{figure}

\begin{figure}
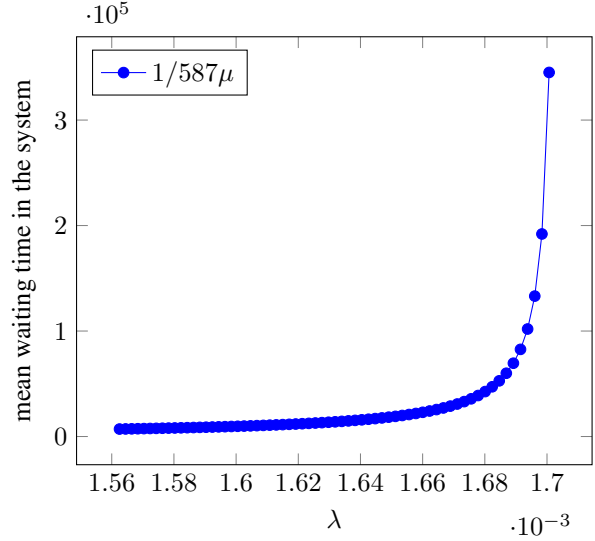

\centering
%
\caption{Mean waiting time in the system}
\label{fig:Mean waiting time in the system}
\end{figure}

\begin{figure}
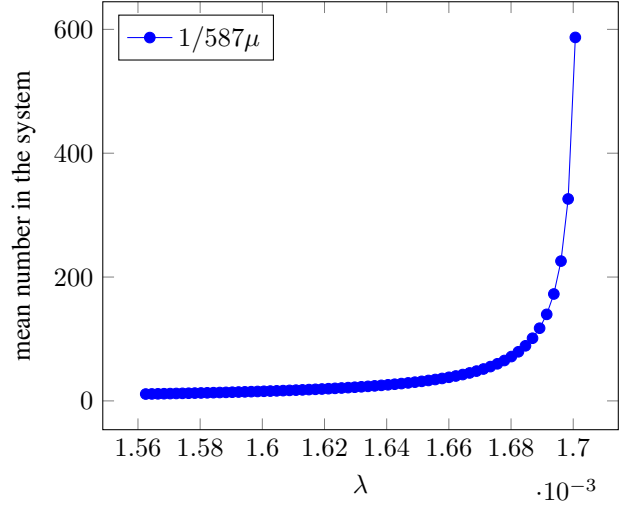

\centering
%
\caption{mean number in the system}
\label{fig:mean number in the system}
\end{figure}

\begin{figure}
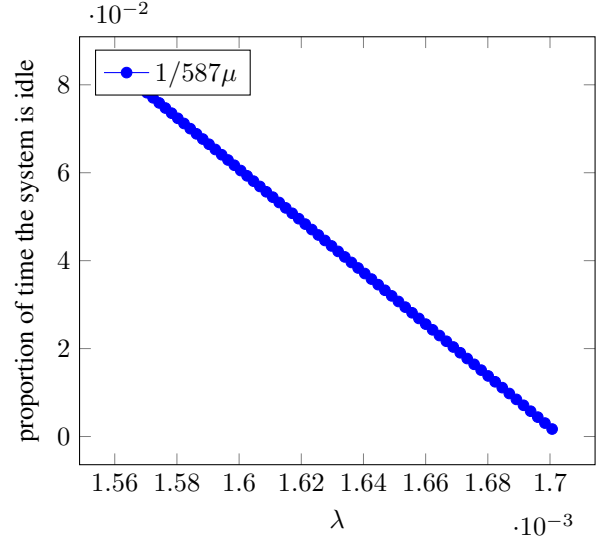

\centering
%
\caption{proportion of time the system is idle}
\label{fig:proportion of time the server is idle}
\end{figure}

\section{Results and Discussion}

\subsection{Results}\label{sec:results}
The system is M/M/1 queuing model with exponential distribution of inter arrival time  and service time.
The system is tested with a mean inter-arrival time between 588 ms and 640 ms.
The mean service rate is $\mu =1/587$, which is the time required for publishing.

The system utilization is computed by Equation  \ref{system utilization} and shown in Figure \ref{fig:System Utilization}. 
Mean waiting number to be served is computed by Equation \ref{Mean waiting number} and shown in Figure \ref{fig:Mean waiting number}.
Mean waiting time is computed by Equation \ref{Mean waiting time} and shown in figure \ref{fig:mean waiting time}. 
Mean waiting number verses mean waiting time is shown by figure \ref{fig:mean waiting time vs  mean waiting number}.

Mean waiting time in the system is computed by Equation \ref{Waiting time in the system} and shown in Figure \ref{fig:Mean waiting time in the system}.
Mean waiting number in the system is computed by Equation \ref{Mean number in the system}  and shown in Figure \ref{fig:mean number in the system}. 
Proportion of time the server is idle is computed by Equation \ref{idle Proportion} and shown in Figure \ref{fig:proportion of time the server is idle}.

\begin{equation}
\label{system utilization}
\rho=\lambda/\mu.
\end{equation}

\begin{equation}
\label{Mean waiting number}
L_q=\frac{\rho ^{2}}{1-\rho}
\end{equation}

\begin{equation}
\label{Mean waiting time}
W_q=L_q/\lambda
\end{equation}

\begin{equation}
\label{Waiting time in the system}
W=W_q+ 1/\mu
\end{equation}

\begin{equation}
\label{Mean number in the system}
L=\lambda W
\end{equation}

\begin{equation}
\label{idle Proportion}
idle\%=1-\rho
\end{equation}

\subsection{Discussion}

As the experiments in section \ref{sec:experiments} and the results in section \ref{sec:results} show, the overhead created between UMA and MQTT is the main critical overhead by the mode.
The UMA and MQTT are originally separate models;
 each model with different features and objectives. 
In the proposed model, MQTT and UMA are integrated to provide extra layer for IoT devices of security.
The main point of adding UMA to MQTT or the vice versa is to add accessibility for users for IoT devices.
Any IoT device is not connected to the world unless connected to a broker or a type of protocol, that allows the IoT device to send and receive data.
However, the accessibility of IoT devices is expected to be quite high; therefore, UMA is provides high scaleability for IoT device to publish its data to the wide range of users or $RP$.
$RP$ could be another IoT device.
Nowadays with the expected future and developing project of smart cities and buildings, IoT devices are expected to send very large amount of data as(i.e. Big Data).
Also, the integration of plug and play is a must future for IoT.
Now, UMA  will establish protocols as well as MQTT.
Adding the two models should be done without sacrificing performance.
If there is a trade-off, the system administrator and project developer can discuss the QoS of any model of expected growing IoT projects.

For MQTT, the model is added to UMA to add the subscriber functionality.
The model shown in Figure \ref{fig:networkmodel}; has mainly $S1$ and$MB2$ added to the $RS$ and $S2$ added to $RP$.
Effectively, $P1$, $MB1$ and $P2$ are not considered an overhead, since it is external processing overhead to the UMA protocol.
Therefore, the overhead are with $S1$, $MB2$ and $S2$.
When the data are received from $P1$ to $MB1$, the data are published to $S1$. $S1$ is represented by $MB2$, which is represented by $RS$.
The integration between $MB2$ and $RS$ is done by considering any newly published data received by $MB2$ as a request received by a $RP$ which is registered as a subscriber for a specific topic.
Therefore, the time required to get a message about a new topic is less than the time required for a request by $RP$
,$Client$
, $S2/RP/xClient$
$ClientxS1/MB2.RS/$
and $S1/MB2/RS/$
compared to the time required for the publishing as mentioned earlier by Equation \ref{Publish}. The time required is only 26\% of the time required if  the functionality is done by UMA only.
Consequently, the performance gain is 74\& with MQTT for the publishing function of a new topic.

\section{Conclusion}
In conclusion, IoT application are promising for future applications in different domains, such as smart cities or the sensing as a service (SenaaS) model of business.
When an IoT application works it is expected that the number of devices grows exponentially as well as the size of data.
Security is required to secure access to the data sources.
Data needs to be protected without compromising performance. 
Therefore, a security layer is required.
In this work, the UMA security protocol is used to add security with the well-known IoT application protocol MQTT.
There are different IoT application protocols; however,MQTT is the most popular for implementations simplicity.
This work has shown how to integrate MQTT and UMA in one fast performance and secure model.
The integration is maintained by mapping the functionality of both protocols.
Performance comparison have shown that there is performance gain for different functionality, mainly the publishing function.
Publishing is the main function of any IoT device.
The model has successfully, connected isolated IoT devices to the publishing network without compromising security.
The data are published and accessed without accessing the source of the data.
UMA is a promising model for IoT security,

%


%
%

\bibliographystyle{unsrt}
\bibliography{references}

\end{document}